\documentclass[final]{epl}
\usepackage{graphicx}
\usepackage{dcolumn}
\usepackage{bm}
\title{Rydberg Atoms in Magnetic Quadrupole Traps}
\author{
Igor Lesanovsky\inst{1}\thanks{\email{ilesanov@physi.uni-heidelberg.de}},
J\"org Schmiedmayer\inst{1}\thanks{\email{joerg.schmiedmayer@physi.uni-heidelberg.de}}, \and
Peter Schmelcher\inst{1,2}\thanks{\email{corresponding author: Peter.Schmelcher@pci.uni-heidelberg.de}}
}
\institute{
    \inst{1} Physikalisches Institut, Universit\"at Heidelberg, Philosophenweg 12, 69120 Heidelberg, Germany\\
    \inst{2} Theoretische Chemie, Institut f\"ur Physikalische Chemie, Universit\"at Heidelberg,INF 229, 69120 Heidelberg, Germany
}
\begin{document}
\maketitle
\begin{abstract}
We investigate the electronic structure and properties of Rydberg atoms exposed to a magnetic quadrupole field.
It is shown that the spatial as well as generalized time reversal symmetries lead to
a two-fold degeneracy of the electronic states in the presence of the external field.
A delicate interplay between the Coulomb and magnetic interactions in the
inhomogeneous field leads to an unusual weak field splitting of the energy levels as well as
complex spatial patterns of the corresponding spin polarization density
of individual Rydberg states. Remarkably the magnetic quadrupole field induces a
permanent electric dipole moment of the atom.
\end{abstract}
The past two decades have seen substantial progress of our knowledge on
highly excited Rydberg atoms exposed to homogeneous magnetic fields providing major
impact on areas such as quantum chaos, semiclassics of nonintegrable systems and properties of
magnetized structures \cite{Friedrich89,Ruder94,Friedrich97,Schmelcher98}.
However, so far there exist no investigations on Rydberg atoms in inhomogeneous and/or trapping magnetic field
configurations. Apart from being of fundamental interest trapped Rydberg atoms have recently been proposed
to serve as a tool for quantum information processing in mesoscopic atomic ensembles
\cite{Lukin01}. For these applications the atoms have to be localized in sufficiently
tight traps providing the appropriate confinement of the highly excited Rydberg states.
Such a tight confinement can be achieved for neutral atoms in magnetic traps on atom chips \cite{Folman02}
where large field gradients ${\mathcal{B}} \approx 10^8 \frac{G}{cm}$ are accessible.
As a prototype example we
study here the magnetic quadrupole field \cite{Bergeman89} that is a key element of magnetic trapping
\footnote{The quadrupole field is not a complete trap in itself}.
We show that Rydberg atoms confined to a quadrupole field possess a specific
structure and symmetry that lead to two-fold degeneracies of its eigenstates. The Rydberg states
exhibit unique phenomena such as complex spin polarization patterns and magnetic field-induced giant electric
dipole moments that can be understood by employing the underlying symmetries and analyzing
the interplay between the Coulomb and magnetic interactions.
We utilize a one-body approach for the Rydberg atom where the motion of the excited outermost
electron takes place in the field of a singly positive charged core. The accuracy of
this assumption increases with increasing degree of excitation and holds particularly well for the
frequently used alkali atoms which possess a single valence electron outside a closed shell core
\footnote{Also, for excited states the spin-orbit and hyperfine interactions can be neglected due
to their rapid drop-off with increasing energetical degree of excitation}.
We assume that the atomic center of mass (CM) is localized at the center of the quadrupole field
requiring an ultracold CM motion of the atom.
The Hamiltonian ${\cal{H}} = {\cal{H}}_{1} + {\cal{H}}_{2}$
in the presence of the quadrupole field $\vec{B}(\vec{r})={\mathcal{B}}(x,y, -2z)$ with the vector potential
$\vec{A}(\vec{r})=\frac{1}{3} [\vec{B}(\vec{r})\times\vec{r}]$ reads
\begin{eqnarray}
{\cal{H}}_1&=&-\frac{\hbar^2}{2m_e}\left(\frac{\partial^2}{\partial
r^2}+\frac{2}{r}\frac{\partial}{\partial r}
+\frac{1}{r^2}\left\{\cot\,\theta\frac{\partial}{\partial
\theta}+\frac{\partial^2}{\partial
\theta^2}\right\} \right. \nonumber \\
&+& \left. \frac{1}{r^2\sin^2\theta}\frac{\partial^2}{\partial\phi^2} \right)
-\frac{1}{4\pi\epsilon_0}\frac{e^2}{r}+i\frac{\hbar e}{m_e}
\mathcal{B}\,r\cos\theta\frac{\partial}{\partial\phi} \nonumber \\
&+& \frac{e^2}{2m_e}\mathcal{B}^2\,r^4 \cos^2 \theta \sin^2 \theta \nonumber \\
{\cal{H}}_2 &=& \mu_B\mathcal{B}r\left(\sin\theta\left\{\sigma_x
\cos \phi+\sigma_y \sin \phi\right\}-2\sigma_z \cos \theta\right)
\end{eqnarray}
We have employed spherical coordinates. $\sigma_i (i=x,y,z)$ are the Pauli matrices acting in spin space.
Here ${\mathcal{B}}$, $\mu_B$ are the field gradient and Bohr magneton, respectively.
${\cal{H}}_1$ contains the Coulomb, paramagnetic and diamagnetic interactions whereas
${\cal{H}}_2$ contains the interaction of the spin with the magnetic field.
Compared to the case of a homogeneous field the Hamiltonian ${\cal{H}}$ exhibits a number of major differences.
Depending on the value of the field gradient ${\mathcal{B}}$ it possesses a strong
variability with respect to the appearance of its energy surfaces. The paramagnetic term
($\propto {\mathcal{B}}$) is, apart from its proportionality with respect to the angular momentum
$L_z=\frac{\hbar}{i}\frac{\partial}{\partial \phi}$, additionally depending on the $z-$coordinate and
the diamagnetic interaction ($\propto {\mathcal{B}}^2$) represents an oscillator coupling term of
fourth order for the motion perpendicular and parallel to the $z-$axis (see ${\cal{H}}_1$).
In addition ${\cal{H}}_2$ results in an intricate coupling
of the spatial and spin electronic degrees of freedom via the inhomogeneity of the field.
It is possible to eliminate the $\phi-$dependence of the Hamiltonian ${\cal{H}}$ by applying the
unitary transformation
\begin{eqnarray}
U=\frac{1}{\sqrt{2}}\left(%
\begin{array}{cc}
  -e^{-i\phi} & e^{-i\phi} \\
  1 & 1 \\
\end{array}%
\right)
\label{3}
\end{eqnarray}\\
acting in spin and angular space. This is due to the
rotational symmetry associated with the conservation of the total angular momentum
$J_z = L_z + S_z$ possessing half-integer eigenvalues $M$. This conservation reflects the
axial symmetry of the quadrupole field. The eigenfunctions of $J_z$ (and ${\cal{H}}$)
take the appearance $|\Phi_m \rangle = \left( \Phi^u(r,\theta) e^{i(m-1)\phi}, \Phi^d(r,\theta) e^{im\phi}\right)$
with $M=m-\frac{1}{2}$, $m$ being integer, and where $\Phi^{u,d}$ are the upper and lower components
of the spinor, respectively.
A close inspection of ${\cal{H}}$ yields a further spatial (unitary) symmetry $P_{\phi}OP_z$ that consists
of the spatial $z-$parity operation $P_z$ followed by an interchange $O(=\sigma_x)$ of the spin components and
the $\phi-$parity $P_{\phi}: \phi \rightarrow 2\pi-\phi$ operation. Additionally ${\cal{H}}$
possesses the generalized time reversal symmetry $TOP_z$ i.e. $\left[TOP_z, {\cal{H}}\right] =0$,
where $T$ is the conventional time reversal operation.
$P_{\phi}OP_z$ and $TOP_z$ do not commute with $J_z$ but yield e.g. $\left[TOP_z,J_z\right]= (-2 J_z) TOP_z$.
Eigenfunctions to the $TOP_z$-operator are provided by the corresponding linear combination
$|\Psi_m^{\pm} \rangle = \frac{1}{\sqrt{2}} \left[ |\Phi_m \rangle \pm TOP_z |\Phi_m \rangle \right]$ with
$TOP_z |\Psi_m^{\pm} \rangle = \pm |\Psi_m^{\pm} \rangle$. Beyond the above we have
the additional symmetry operation $TP_{\phi}$ that commutes with both the Hamiltonian ${\cal{H}}$
and the angular momentum $J_z$.
Employing $\left\{TOP_z,J_z\right\}=0$ it can be shown that each energy eigenvalue is doubly degenerate
with the two energy eigenstates being
$|\Phi_m\rangle, |\Phi_m\rangle^{\prime} = TOP_z |\Phi_m\rangle$
i.e. they possess the eigenvalues $\pm M$ with respect to $J_z$.
The underlying symmetry group is Nonabelian and a semi-direct product
$C_{\infty v} = C_{\infty} \bigotimes C_{s}$. This is in contrast to the case of a homogeneous magnetic
field where $L_z, P_z, P, T \sigma_z P_{\phi}$ constitute the spatial and time reversal symmetries,
respectively, and form
an Abelian group thereby not causing any degeneracies due to symmetry.
The above symmetries and degeneracies have to be carefully distinguished from the two-fold Kramers
degeneracy of spin $\frac{1}{2}$ systems in the absence of the field. The latter is lifted if
an external (even homogeneous) field is switched on. The degeneracies found here are due to the
particular geometry of the quadrupole magnetic field.
To investigate the electronic structure of the atom exposed to the field in detail we expand the eigenfunctions
of ${\cal{H}}$ in two-component spinors according to
\begin{equation}
\Psi(r,\theta,\phi) = \sum \limits_{n,l,\tilde{n},\tilde{l}} c_{n,l,\tilde{n},\tilde{l}}
\left( \begin{array}{c} R_{n}^{(\zeta,k)}(r) Y_{l}^{(m-1)} (\theta,\phi)\\
R_{\tilde{n}}^{(\zeta,k)}(r) Y_{\tilde{l}}^{(m)} (\theta,\phi)\\
\end{array}
\right)
\end{equation}
where $(r,\theta,\phi)$, $Y_{l}^{(m)}$ are spherical coordinates and the spherical harmonics, respectively,
and $R_{n}^{(\zeta,k)}(r)=\sqrt{\frac{n!}{(n+2k)!}}e^{-\frac{\zeta r}{2}}(\zeta r)^{k}L_n^{2k}(\zeta r)$
with $L_n^{a}$ being the associated Laguerre polynomials. $\zeta$ is a nonlinear variational parameter
to be optimized. We apply the energy variational principle linearly optimizing the coefficients
$c_{n,l,\tilde{n},\tilde{l}}$ by solving the corresponding generalized eigenvalue problem (GEP). The Hamiltonian
and overlap matrices can be calculated analytically possessing a band and block structure, respectively,
that can be exploited in solving the GEP. Our approach to diagonalize the GEP consists of a Krylov-space approach
using the Arnoldi-decomposition and furthermore applying a shift-and-invert procedure \cite{Sorensen,Demmel91}.
This allows us to accurately describe highly excited states since the shift-and-invert approach
combined with an optimized value for the parameter $\zeta$ ($1/\sqrt{0.775 |E|}$ is a good choice)
allows to converge the eigenvalues in a preselected window of the excitation energy.
For the field gradients up to ${\mathcal{B}} = 10^{13} \frac{G}{cm}$ we have studied excitation spectra
corresponding to the magnetic quantum numbers $M = (-\frac{7}{2}) - (+\frac{7}{2})$.
For the gradients presently accessible for atom chips
(${\mathcal{B}} \le 10 ^{7} \frac{G}{cm}$) all Rydberg states up to $n \approx 60$ ($n$ refers to the
corresponding field-free principal quantum number which is not a good quantum number in the presence
of the field but serves as an energetical label) could be calculated with high accuracy.
To investigate the properties of the excited states in more detail we analyzed the spectrum $E_i({\mathcal{B}})$
($i$-th energy curve), the spatial probability densities $W_{\Lambda} (r, \theta)  = r^2 \sin \theta |\Phi_m(r,\theta)|^2$
of individual states, the $z$-component of the spatial spin density $W_S (r,\theta) = \frac{\hbar}{2}
\frac{|\Phi^u(r,\theta)|^2-|\Phi^d(r,\theta)|^2|}{|\Phi^u(r,\theta)|^2+|\Phi^d(r,\theta)|^2|}$
as well as several relevant expectation values such as dipole moments and spin polarizations.
\begin{figure}
\oneimage[width=14cm]{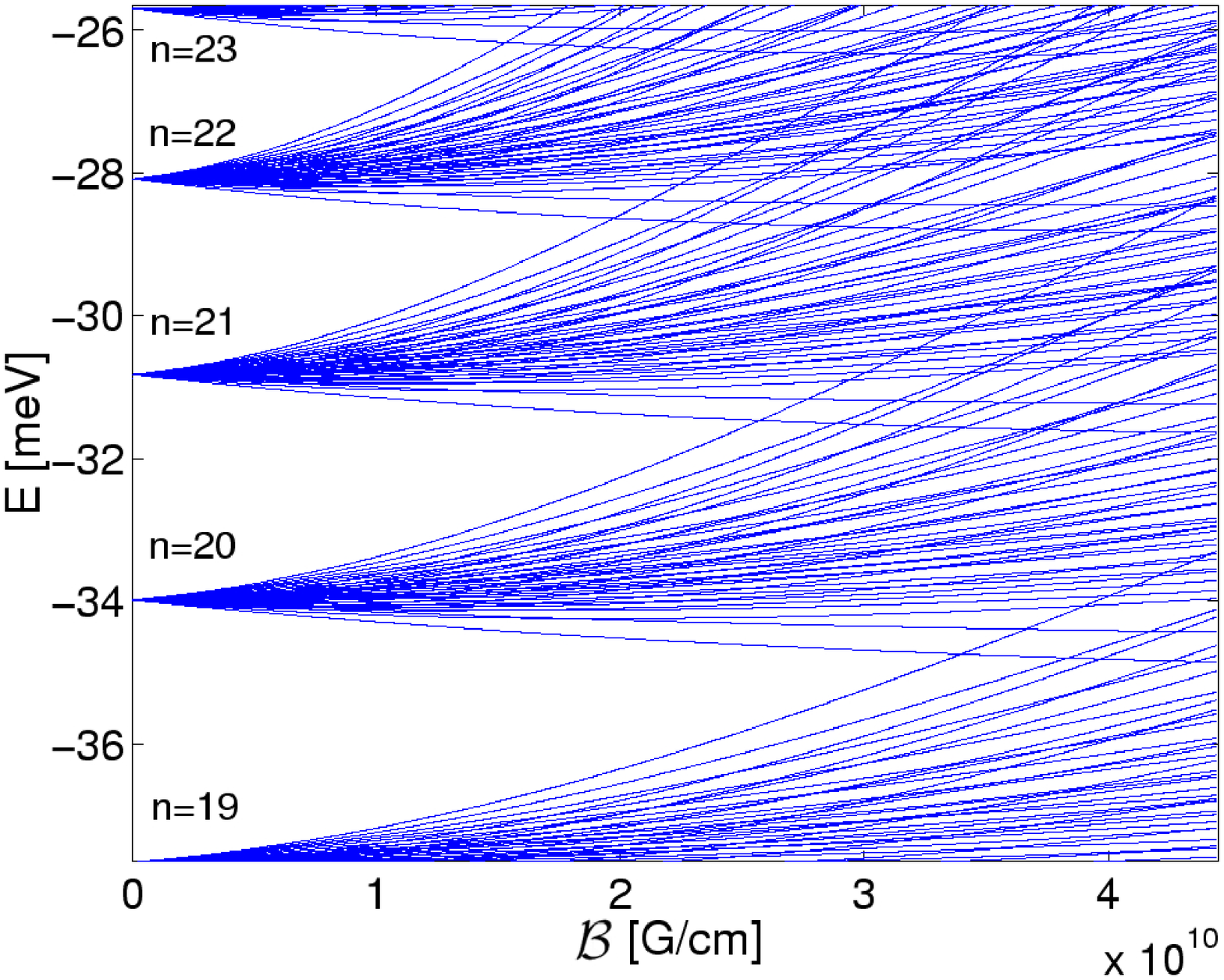}
\twoimages[width=7cm]{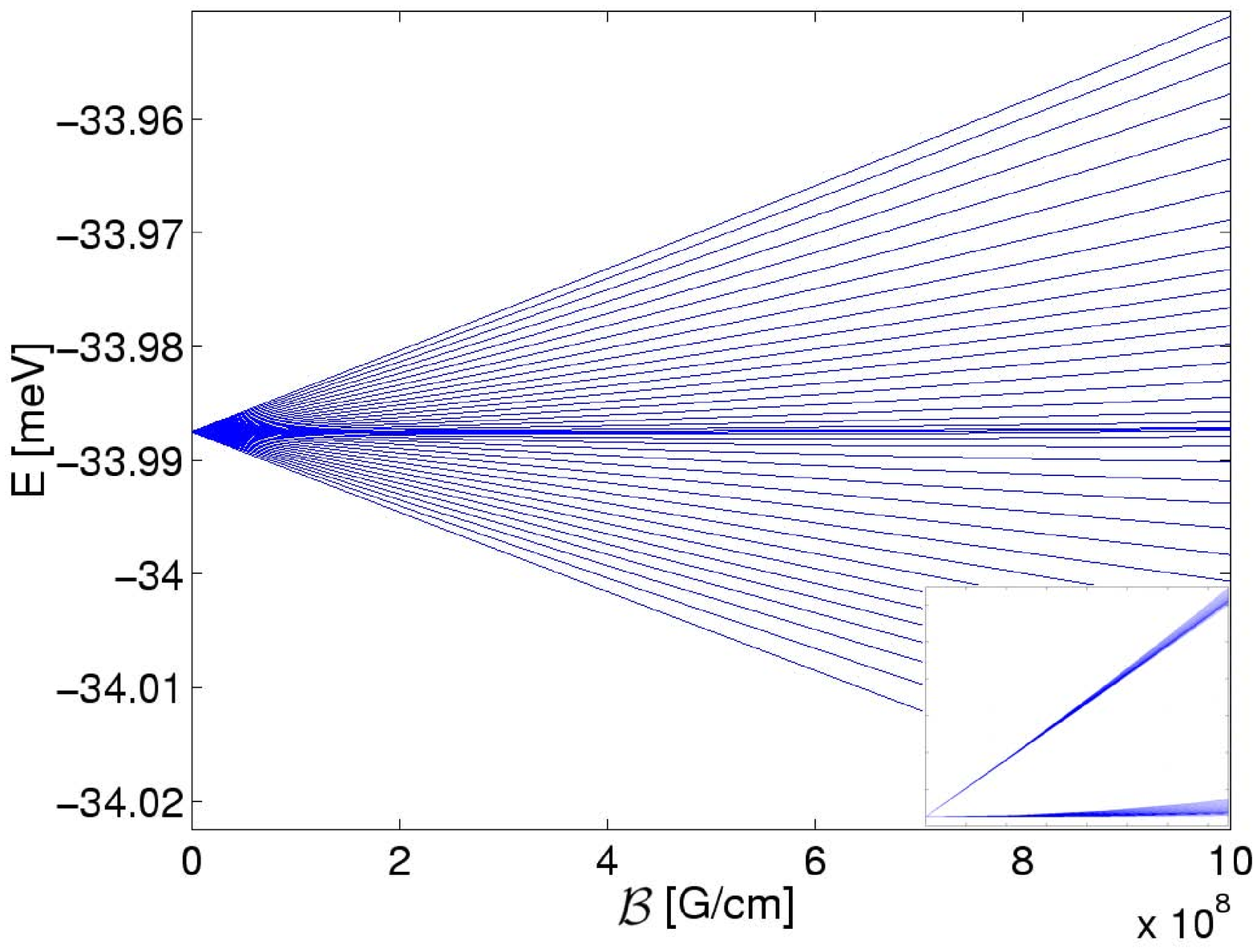}{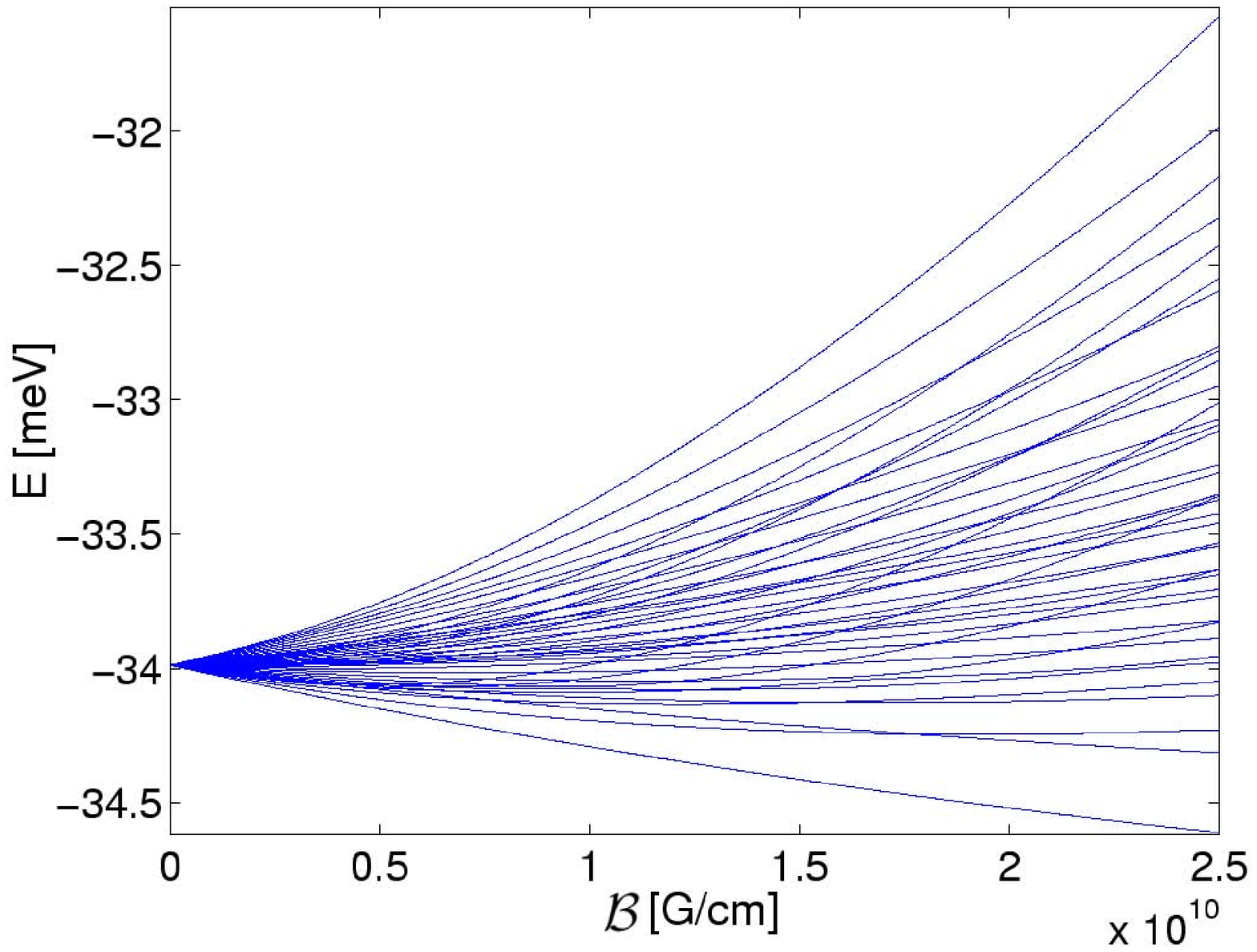}
\caption{Upper panel: The energy levels as a function of the field
gradient ${\mathcal{B}}$ emerging for the $n=19-23, M=\frac{1}{2}$
manifolds. The inter $n$-manifold mixing for strong gradients is
clearly visible. Lower right panel: Symmetric linear splitting of
the $39$ (for ${\mathcal{B}}=0$) degenerate energy levels
belonging to the multiplett $n=20, m=1$ for weak gradients. Inset:
The splitting of the same states in a homogeneous magnetic field
covering the range $0 \le B \le 0.235$ Tesla. Lower right panel:
The splitting of the same multiplett for a larger range of field
gradients covering the regime of $l$-mixing. Very narrow avoided
crossings occur in this regime. }
\end{figure}
Emerging from ${\mathcal{B}}=0$ we encounter for weak gradients
a splitting of the degenerate energy levels that looks
quite different from what is observed in the case of a homogeneous magnetic field. Focusing e.g.
on $M=\frac{1}{2}$ and a given $n-$manifold the splitting in a homogeneous field takes place into
two bundles each consisting of almost degenerate $n$ and $n-1$ sublevels (see inset of upper panel of
figure 1), respectively. One of the bundles is (approximately) independent of the field and the
other one raises linearly with increasing field strength. In the quadrupole field the degenerate
$n-$manifold of $M=\frac{1}{2}-$states splits into its $2n-1$ components i.e. energy curves
$E_i({\mathcal{B}})$ (see upper panel of figure 1).
Each curve behaves approximately linear with increasing field gradient
(in the weak gradient regime where the paramagnetic interaction dominates) but possesses a different slope.
The curves $E_i({\mathcal{B}})$ are arranged symmetrically with respect to the constant $E_i(0)$ as
can be seen in figure 1. With increasing ${\mathcal{B}}$ the clusters of levels widen further and the curves
$E_i({\mathcal{B}})$ become nonlinear (see middle panel of figure 1).
The field gradient for intra $n$-manifold mixing of different
angular momentum i.e. $l$-states can be shown to scale as ${\mathcal{B}} \propto n^{-6}$.
In this regime very narrow avoided crossings occur.
With further increasing field gradient the $n$-manifolds start to overlap
(see lower panel of figure 1) and we encounter inter $n$-manifold
mixing that scales according to ${\mathcal{B}} \propto n^{-\frac{11}{2}}$ (in a homogeneous field
the corresponding scaling for inter $n$-manifold mixing is $B \propto n^{-\frac{7}{2}}$).
In this regime the diamagnetic interaction of ${\cal{H}}_1$ is important and no (not even approximate) symmetries
remain. Level repulsion and avoided crossings are therefore a characteristic feature of the spectrum.
\begin{figure}
\onefigure[width=7cm]{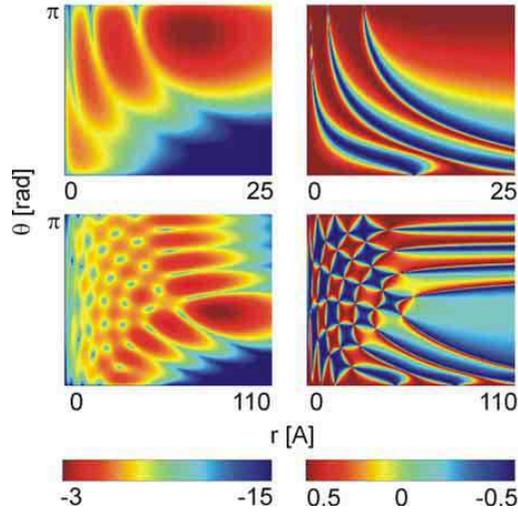} \caption{The spatial
probability density $W_{\Lambda}$ (left column, logarithmic
representation) together with the $S_z$-density in
$r,\theta$-space (right column) for the $15$-th excited state with
$M=\frac{1}{2}$ for ${\mathcal{B}}=4.4 \cdot 10^{9} \frac{G}{cm}$
(upper panel) and for the $76-$th excited state (lower panel)
emerging from $n=9$ with $M=\frac{1}{2}$ for the same gradient.
The asymmetry of the probability density with respect to the line
$\theta= \frac{\pi}{2}$, i.e. for $z \rightarrow -z$, is clearly
visible.}
\end{figure}
In the quadrupole field the spatial probability distributions of the $J_z-$eigenstates
are typically localized in one half-volume either above or below the $x,y-$plane
(see left column of figure 2 which shows the distributions
for the $15$-th and $76-$th excited state for ${\mathcal{B}}=4.4 \cdot 10^{9} \frac{G}{cm}$)
\footnote{This property is related to the existence of permanent electric dipole moments as discussed below}.
With increasing degree of excitation of the eigenstates, the diamagnetic interaction becomes important
and leads to an additional deformation of the electronic probability density. In a homogeneous magnetic field
$z$-parity is a symmetry and $W_{\Lambda}$ is symmetric with respect to reflections at the $x,y$-plane.
Considering the expectation value $\langle r \rangle$ for the states with increasing $n$ the effect of the quadrupole
field is twofold: it modifies the distribution of $\langle r \rangle$ values within a single $n-$manifold and
decreases the center of the distribution due to the compression of the electronic cloud in the quadrupole field.
Looking at the spatial $W_S$-spin density of the electronic states some remarkable properties appear.
Figure 2 shows $W_S(r,\theta)$ for the $15$-th and $76-$th excited state for a field gradient
${\mathcal{B}}=4.4 \cdot 10^{9} \frac{G}{cm}$. For the $15$-th excited state $W_S$ shows curved
stripes of upward and downward polarized spin, respectively, that match with the regions
of the localization of the spatial probability density $W_{\Lambda}$.
For the $76-$th excited state a pattern of nested islands appears with each island possessing a certain spin
polarization and well-localized transition regions separating them.
These islands correspond to locally either upward or downward
pointing spin and are arranged in a chess board-like pattern. The borderlines between the islands
correspond to a vanishing $z$-component of the spin.
The intersection of the borderlines i.e. the corners of the islands, represent the nodes
of the spatial probability densities of the Rydberg states as can be seen from the corresponding graph $W_{\Lambda}$
in figure 2.
With increasing value of $r$ the shape of the islands become elongated in radial direction
and finally turn into directed stripes with continuous transitions of the spin polarization.
The formation of the islands is due to a detailed balance of the interactions in the Hamiltonian ${\cal{H}}$.
This does not occur for the atom in a homogeneous magnetic field (constant spin polarization)
nor for the case of a pure spin in a quadrupole magnetic field described by ${\cal{H}}_2$ only.
The latter yields a spin polarization density that is independent of $r$
showing horizontal stripes in the $(r,\theta)$-representation of $W_S$.
These stripes can be found for $W_S$ of the Rydberg states
in figure 2 for large values of $r$ in a modified form indicating the dominance of the
spin coupling to the magnetic field in this regime. Analyzing the spin density
of many states we found that the above behaviour is generic.
The spin polarization patterns are features of individual electronic eigenstates and become
increasingly more detailed with increasing degree of excitation of the state considered.
\begin{figure}
\onefigure[width=7cm]{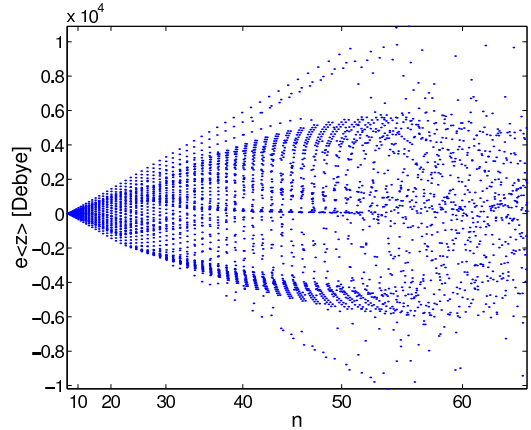} \caption{The expectation
value of the electric dipole moment along the $z-$axis for
${\mathcal{B}}=4.4 \cdot 10^{8} \frac{G}{cm}$ for the states with
$M=\frac{1}{2}$. $n$ labels the excitation energy.}
\end{figure}
Transitions among eigenstates to the total angular momentum $J_z$ obey the selection rules $\Delta M = 0$ and
$\Delta M = \pm 1$ for dipole transitions via linear and circular polarized light, as it is the case
without the presence of the field. Linearly $z-$polarized transitions between $TOP_z$-eigenstates
must involve states with different $TOP_z$ symmetries in order to possess a nonvanishing dipole strength.
The fact that the quadrupole field causes a nonsymmetric charge distribution with respect to the horizontal
$x,y-$plane leads to the following peculiar properties of the atom:
Electronic eigenstates to $J_z$ possess a nonvanishing permanent electric dipole moment $e \langle z \rangle$
only along the symmetry axis of the quadrupole field i.e. the external magnetic field induces a
permanent electric dipole moment of the atom. Figure 3 illustrates the distribution of dipole moments
for a variety of states belonging to the manifolds $n \le 65$. The variance of the distribution of
dipole moments increases strongly with increasing degree of excitation $n$ thereby showing a transition
from a regular alignment to an irregular spreading. In a homogeneous magnetic field the
deformation of the charge distribution is (due to parity symmetry) such that the electric
dipole moment vanishes.
One possibility to probe the above-described properties of Rydberg atoms
in quadrupole magnetic fields would be to perform spectroscopy of single atoms in traps on, preferably, atom
chips since these possess currently the strongest available field gradients. This would provide
us with detailed information on the level splitting and evolution with increasing degree of
excitation.

Already in the presence of a homogeneous magnetic
field it is well-known that the CM and electronic motion of an atom do not decouple
\cite{Avron78,Johnson83,Schmelcher94}. To enter the corresponding regime where the residual coupling
becomes important certain parameter values (excitation energy, CM energy etc.)
have to be addressed. A variety of intriguing phenomena due to the
mixing of the internal and CM motion such as the diffusion of the CM or giant dipole states
are then observed \cite{Schmelcher92,Dippel94,Schmelcher95}.
In the quadrupole field the assumption that the atom is ultracold will certainly
minimize the CM motional effects. Nevertheless, a residual coupling is unavoidable and its impact
on the electronic structure is, at this point, simply unkown: A full treatment of the two-body
system certainly goes beyond the scope of the present investigation and requires both from the
conceptual as well as computational point of view major investigations. On the other hand one should note
that the symmetries discussed here equally hold for the moving two-body system i.e. the total
angular momentum is conserved and the unitary as well as antiunitary spin-spatial symmetries, now applied
to both particles, are also present.

Beyond the above, one can speculate about potential applications of the magnetic field-induced
permanent electric dipole moments of the atoms. Populating with a laser excited states with
a desired dipole moment for certain atoms within an array of single atom traps
can open the route to a controlled interaction between the atoms
which is currently of major interest for quantum information processing \cite{Jaksch99,Calarco00,Lukin01,Eckert02}.
Discussions with Ofir Alon and M. Anderson are gratefully acknowledged. J.S. acknowledges financial support
by the European Union contract numbers IST-2001-38863 (ACQP).

\end{document}